# An Overview of Microinverter Design Characteristics and MPPT Control

Sean Ritson
School of Electronics, Electrical Engineering and Computer Science (EEECS), Queen's University Belfast, BT9 5AH, UK
s.ritson01@qub.ac.uk

Ahmad Elkhateb
School of Electronics, Electrical Engineering and Computer Science (EEECS), Queen's University Belfast, BT9 5AH, UK
A.Elkhateb@qub.ac.uk

**Abstract** - Micro-inverter technologies are becoming increasingly popular as a choice of grid connection for small-scale photovoltaic systems. Efficiently harvesting the maximum energy from a photovoltaic system reduces the Levelized cost for solar energy, enhancing its role in combatting climate change. Various topologies are proposed through research and have been summarised in this paper. Furthermore, this paper investigates two popular Maximum Power Point Tracking (MPPT) methods through simulation using Matlab Simulink.

## I. Introduction

Traditionally, our power system has been a passive network, whereby the bulk supply of electricity provided to the grid has been provided by a small number of large-scale generation sources such as large thermal power plants. However, in recent years we have witnessed an intense uptake in the amount of renewable energy generation embedded within our power system. The growing penetration of renewable generation is an effort to tackle ongoing concerns regarding climate change and meet increasing demand from the electrification of essential industries such as heat and transport. The most popular of these renewable energies include wind, solar, hydro, and biomass, and these sources can connect at the transmission or distribution level.

Solar power is prominent in providing electricity to support industries, transport, and everyday consumers. Solar power extracts energy from solar irradiance and converts it to electrical energy using Photovoltaic (PV) modules and DC-DC and DC-AC converters [1-4]. From 2007 to 2018, the world's total PV capacity increased by nearly 4,400%, from 9.2GW to 404.5GW [5]. In 2018 PV generation accounted for 29.6% of the overall installed renewable penetration within the UK [5]. The significant growth in installed PV capacity presents an opportunity to improve on the performance and availability of the technology.

The electricity generated by PV modules is extracted and converted to Alternating Current (AC) using inverter technologies. In the past, PV arrays consisted of modules connected in series and parallel configurations, which coupled to a central inverter to regulate the output to the grid or consumer. Centralised or string inverters are deployed for harvesting energy from more extensive connections, generally, anything rated above a kilowatt (kW). The issue regarding the traditional methods of connection is partial shading or temperature fluctuations impacting on individual modules in an array can considerably reduce the power output, and therefore reduce the efficiency of the entire system. Therefore, issues occur when installing small-scale PV systems like those on houses, schools, factories, or small businesses where partial shading from chimneys, trees, or surrounding buildings may be a regular occurrence. The primary solution to improve the efficiency of small-scale PV systems is the micro-inverter.

Micro-inverters are connected to individual PV modules and are required to be small devices, to reduce the heat expanded onto the module and fit within a confined space. The general functionality of a micro-inverter is to step-up the voltage from the module and convert the output to a sinusoidal waveform. The step-up converter is required to increase the low output voltage of the PV module, which is typically around 15 - 40V for a standard module. Micro-inverters typically employ conventional DC-DC converters or transformer topologies to increase the low PV voltage. The conversion from DC to AC commonly uses a DC-AC inverter. Figure 1 below shows the typical configuration for converting solar irradiation into usable electricity.

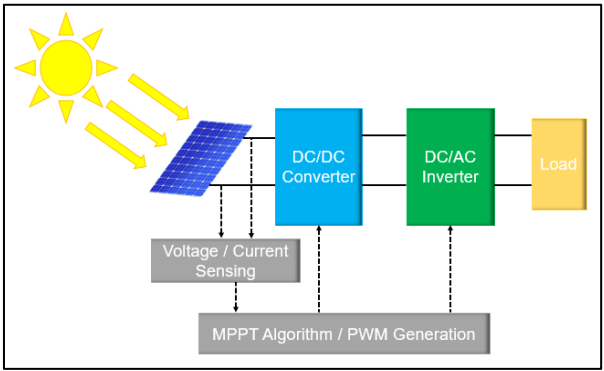

*Figure 1 -Typical Micro-inverter Topology*

Initially, this paper provides an overview of the grid-connection strategies for the standard solar inverter. Next, a literature review analyses the popular micro-inverter topologies and industry research. An introduction to MPPT algorithms is provided through the description and simulation, which implements two popular MPPT algorithms.

## II. Inverter Connection

Inverters are required to harvest Direct Current (DC) electrical energy from PV modules and convert it to Alternating Current (AC). There are a variety of options when it comes to connecting the PV modules to inverters, and the configuration required is specific to the application. The most common configurations are the centralised inverter, the string inverter and the micro-inverter.

The centralised inverter topology shown in Figure 2 below shows three strings of series-connected PV modules which are then connected in parallel and feed into a single inverter. The outputs from the individual strings connect through a combiner before being inverted to AC. This configuration is regularly used for high power connections where the connection is rated from several kW to the MW range.

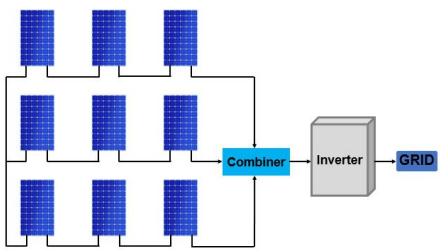

*Figure 2 - Centralised Inverter*

This configuration is favourable due to the ease of its design and implementation. The wiring for the configuration is simplistic compared to the other topologies and thus makes maintenance routine. Additionally, the use of only a single inverter can reduce the capital cost of the system.

The disadvantages of the design include there is only a single MPPT implementation for the whole array. Thus, partial obstruction or blocking of sunlight over a single module will reduce the power output from the system.

When connecting PV modules using this arrangement, the wiring used in the lead up to the converter contains DC, which incurs more significant power losses over long distances [6]. The centralised inverter configuration carries the risk of a single failure to the inverter, causing failure of the entire system.

Figure 4 shows the typical configuration for a string inverter. The arrangement shows that for the same number of PV modules as the centralised inverter there are three inverters required. Each string of PV modules consists of multiple modules cascaded in series feeding into an inverter. This arrangement would commonly be used for systems rated from 500W to several kW.

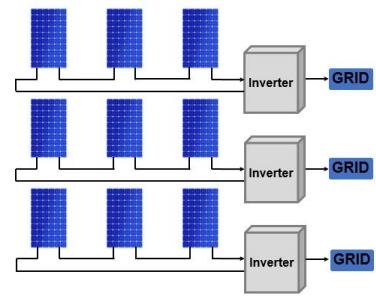

*Figure 3 - String Inverter*

The primary advantage is the implementation of MPPT control for each string of PV modules, thereby increasing the power delivered and improving the overall efficiency of the system compared to the centralised inverter. Also, the reduced length of wires required to carry DC reduces the connection losses. The design of the system is an open design that can be expanded through the addition of more strings.

Similar to the centralised inverter, the impact of partial shading on a single PV module will significantly inhibit the overall power output that can be supplied by the string inverter. Partial shading reduces the Maximum Power Point (MPP) of the shaded module, and as the modules are connected in series, the string MPP will be limited to the lowest panels MPP. The requirement for more inverters can also incur higher capital expenditure for the system.

The final configuration to discuss is the micro-inverter topology which is the focus of this paper. The configuration for a micro-inverter connected system is shown in Figure 5 below. The micro-inverter employs a single inverter for each PV module, thereby providing increased control capability and fault resilience. Micro-inverters are typically deployed for systems where each PV module is rated up to 500W.

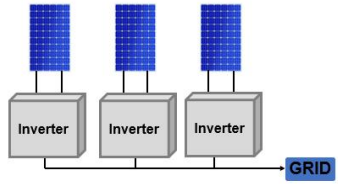

*Figure 4 - Micro-inverter*

The main advantage of using micro-inverters is the increased MPPT coverage which allows the individual modules to operate independently. Therefore, if there is partial shading on one of the modules, it will not negatively impact the surrounding unshaded modules power output. Therefore, under partial shading conditions, the micro-inverter system will be more efficient when compared to the centralised or string inverter systems.

The micro-inverter removes the mismatch between PV module voltage, which appears in the centralised and string inverter arrangements, thereby maximising the output power available and improving system efficiency [6]. The operation of each module independently means the system can be easily extended with the inclusion of additional modules and inverters without any redesign to the current system. Previously discussed was the DC wiring required for the centralised and string inverters; however, as micro-inverters can reside close to the PV module this wiring is reduced, and consequently, DC losses are reduced.

The drawback of this configuration is the requirement for additional inverters. Therefore, this arrangement is generally avoided for large systems due to the significant capital cost. As previously mentioned, the micro-inverters reside close to the modules and whilst this can reduce losses, it can increase the maintenance complexity as the inverters can be difficult to access.

### III. Inverter Topologies

A single-stage micro-inverter topology is presented in [7] which combines a flyback and buck-boost converter utilising a high-frequency flyback transformer. The equivalent circuit is shown in Figure 6 below. The primary side includes the addition of a buck-boost converter to the standard flyback converter topology providing an auxiliary power decoupling strategy [8]. Therefore, the decoupling capacitor CPV can be drastically decreased as much as 1/100th or 1/200th of the equivalent capacitor required if the buck-boost element was removed [7]. Alternatively, a large PV side decoupling capacitor would be required to filter out the double-line-frequency ripples [9].

One issue associated with this topology is the overvoltage which can be present across SDC during turn off as a result of leakage inductance in the flyback transformer [9-11]. Leakage inductance is a practical consideration in transformers due to imperfections between windings. It can result in losses across the transformer, consequently reducing the converter efficiency. This makes the flyback topology unattractive when considering applications over 200W [12].

Typical flyback converters utilise a clamp or snubber with a complimentary gate signal to eliminate this [11]. However, this strategy cannot be used alongside flyback converters which have a variable switching frequency [11]. Furthermore, [10] presents that if an RCD clamp is used in this case the power losses in the RCD clamp can reduce overall efficiency or the interaction between the RCD clamp and buck-boost converter can cause the inverter to fail.

A non-complimentary active clamp can be utilised for the flyback converter to reduce leakage inductance losses, thereby improving efficiency [9], [11]. Alternatively, the converter can be replaced with a dual transistor flyback converter, which can achieve a European Efficiency of 81.8% for a 160W inverter as presented in [10].

To investigate the potential for using high-gain converters, various boost converter topologies were reviewed. Numerous topologies implement Uninterruptable Power Supply (UPS) configurations whereby a boost converter is deployed to step up DC input voltage. This is essential in grid-connected micro-inverter applications where the voltage is required to be stepped up from around 15-40V for a single module [13].

A drawback of the conventional boost converter is at significantly low DC input voltages, an exceptionally large duty cycle (approaching 100%) is required to achieve the step-up gain, discounting the performance of the inverter [14].

The performance of the conventional converter topology is limited due to the large current ripples of the switches and output diodes. Also, the switching and diode reverse-recovery losses are significant due to hard switching and high switch voltage stress [15]. It is widely recognised that reducing switching losses can result in considerable efficiency gains.

A double boost configuration is presented whereby the inductors are replaced by coupled inductors [16]. This enables the converter gain to be extended by the configuration of the turns ratio. Furthermore, following experimentation [16] declares the configuration can exert 97.5% efficiency whilst maintaining Total Harmonic Distortion (THD) under 3%.

[17] presents a multiple-input boost converter which aims to execute MPPT for individual strings of cells within a PV module. This further reduces the impact of partial shading. The analysis and simulation proposed concluded that a 9.6% improvement in efficiency is possible using this strategy. However, the more sophisticated hardware configuration will increase the overall design cost.

In cases, resonant converter schemes are used within micro-inverter topologies. These schemes are appealing as they enable high switching frequencies to achieve compact converters whilst maintaining a high efficiency [18]. An Interleaved Isolated Boost Series Resonant Converter (IBSRC) utilises this scheme claiming fast rejection of disturbances during input and output variations [12].

An isolated interleaved boost converter is presented in [19] which states wide-bandgap SiC MOSFETs are advantageous due to their high switching frequency and high thermal conductivity making them highly efficient yet also being small in size. Wide bandgap semiconductor devices have been researched extensively to understand the potential to be used in converter applications [20]. Due to the device's small ON-state resistance, the overall switching losses are reduced, improving efficiency [19].

## IV. Power Decoupling

Power Decoupling is an important aspect when designing a micro-inverter. The decoupling capacitor's role is to suppress unwanted components of the supply signal. Three power decoupling techniques are deployed across micro-inverters including PV Side Decoupling, DC Link Decoupling and AC Side Decoupling. In general, PV side decoupling requires the largest capacitance value [21].

The inclusion of large electrolytic capacitors increases the size of the micro-inverter, as well as adding to the weight. Using bulky capacitors will negatively impact on the lifetime of the micro-inverter and will require replacement before other components, including the PV module itself. Furthermore, large decoupling capacitors can be prone to failure under adverse conditions, such as high temperature [22], reducing the lifetime of the design. When implementing a pseudo-DC link, which is essentially a DC rectified AC inverter followed by an Unfolder, it is common to use a decoupling capacitance on the PV side.

*A. PV Side Decoupling*

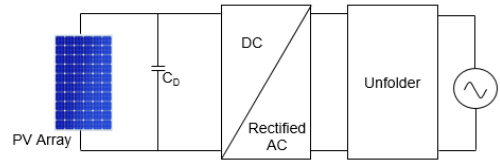

*Figure 5 - PV Side Decoupling*

[23] provides a sample calculation for a standard PV panel calculating the required value of decoupling capacitance using the following equation:

$$C = \frac{P_{PV}}{\omega_0 V_{DC} \Delta V} \quad (1)$$

Using $\omega_0 = 2\pi f_0$ with a fundamental frequency of 50Hz. The power rating of the PV panel PPV = 200W. The link voltage VDC = 35V. The voltage ripple $\Delta V$ = 2V. The calculated value for decoupling capacitance CD is 9.09mF. This represents a large value of capacitance which means the micro-inverter will be physically larger. As previously mentioned, using a bulkier electrolytic capacitor will also negatively impact on the lifetime of the micro-inverter.

It is possible through the introduction of more complex circuitry to reduce the value of decoupling capacitance required on the PV side, such as described previously for the flyback converter topology. It is generally undesirable to introduce more complex circuitry unnecessarily as this can reduce the power density and lifetime of the design as well as incurring additional cost.

*B. DC Link Decoupling*

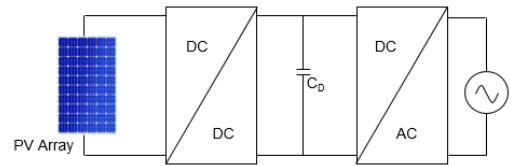

*Figure 6 - DC Link Decoupling*

It is proposed in [23] that for the same circuit parameters, using the DC link decoupling strategy reduces the size of the capacitor required to 1/70th the size of the equivalent electrolytic capacitor required for PV Side decoupling. Furthermore, [21] discusses that implementing the DC link decoupling method allows for a higher DC link voltage as well as a higher ripple voltage. [23] supports the potential for a significant reduction in capacitance value when moving from PV side decoupling to DC link decoupling. On top of this, there is no additional circuitry or complex control required, which both simplifies the design and reduces cost. For this reason, the DC link decoupling technique is commonly used across micro-inverter designs when reducing the size is of significant importance.

*C. AC Side Decoupling*

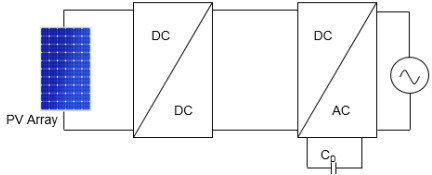

*Figure 7 - AC Side Decoupling*

The last method of power decoupling to discuss is the AC side decoupling. It is suggested that this method can dramatically reduce the required capacitance. [21] outlines the challenges incurred with this approach, starting with the decoupling capacitor usually needs to be embedded into the inverter stage introducing an additional phase leg. This can reduce the overall efficiency whilst introducing additional control, thus increasing the cost.

## V. MPPT Operation

One of the critical research areas in the development of micro-inverters is the potential to better utilise the converter control schemes to improve device efficiency. The most common control implementation is using Maximum Power Point Tracking (MPPT). This involves sensing the current and voltage output from the PV module and adjusting the duty cycle of the switching device (normally MOSFETs or IGBTs) to ensure the maximum power is being extracted from the PV module at any time.

Obtaining maximum power from the module is a crucial factor in maximising the energy harvest and efficiency of the PV system. The maximum power obtainable from a PV module can vary according to alterations in the following key factors:

- Irradiation – during periods of greater irradiation intensity, the maximum power available from the PV module increases. The impact irradiance has on the MPP of a standard module rated at 213W is shown in Figure 8 below.

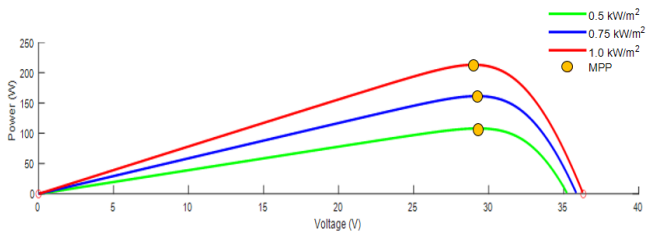

*Figure 8 - Impact of Irradiation on MPP*

- Sunlight incidence angle – the maximum power available from a module is at its highest when the photons arriving on the surface of the module are arriving perpendicular to the surface

- Temperature – in hotter conditions, the maximum power from PV modules is decreased. The temperature effect of the module can be a result of an increase in the ambient temperature or the solar cell's internal temperature. The impact temperature has on the MPP of a module is shown in Figure 9 below.

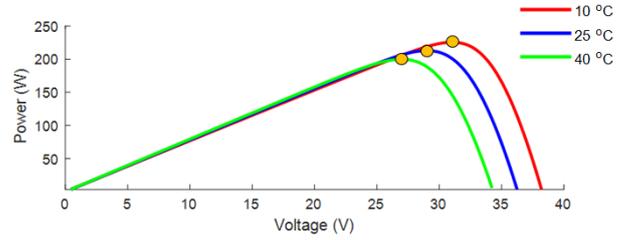

*Figure 9 - Impact of Temperature on MPP*

The direct MPPT methods include P&O, IC, Curve Fitting and Fuzzy Logic. Each MPPT algorithm operates with unique characteristics that can be analysed in terms of response, stability and complexity.

The Curve Fitting method uses pre-specified PV characteristics alongside a defined mathematical formula to accurately predict the MPP. Using an $n^{th}$ order polynomial, the voltage at which the power is at its maximum for a given set of conditions can be obtained. The advantage is this method can easily be implemented; however, it requires extensive prior knowledge of the specific PV module characteristics and cannot be easily adapted to alternative systems. Additionally, the calculations use a significant amount of memory [24].

Research has presented limitations with the traditional methods of control, such as when the system is subject to sudden variations in input conditions. A Fuzzy Logic controller is presented in [25], which aims to alleviate some of the restrictions on traditional methods. The proposed strategy provides a fast response time reducing variations in voltage, power and duty cycle. The controller also reduces the efficiency losses caused by oscillations.

Also available is Sliding Mode control which provides enhanced stability and accuracy over linear methods such as P&O [26].

The P&O algorithm is a commonly used MPPT algorithm. A typical functional flow diagram is shown below in Figure 10.

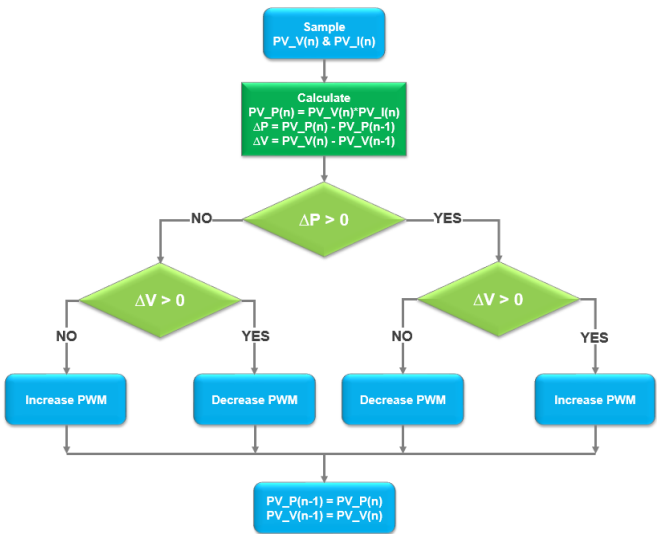

*Figure 10 - P&O Functional Diagram*

The P&O algorithm is considered to function similarly to hill-climbing control methods. P&O and hill-climbing methods operate on the foundation that the power or voltage becomes zero at the maximum power point [27] and the duty cycle can be increased or decreased accordingly to control the approach to the MPP.

The P&O is widely used due to its simplicity in implementation. The method also presents greater applicability to on-grid PV systems due to having a well-regulated output and fast-acting dynamic response [28-30].

There are some limitations with the algorithm such as the technique uses a fixed step increase, and this has the potential to introduce steady-state oscillations during operation which can also result in a tracking delay [31]. Choosing a smaller step increase will reduce steady state oscillations at the expense of slower response speed of the MPPT algorithm and choosing a larger step increase will have the reverse effect [32].

A solution to the steady-state oscillations issue is presented in [31] which introduces a Modulated P&O algorithm whereby the step increase is variable. This operates through minimising the step increase during steady-state conditions and increasing the step increase during periods where greater change is required.

The Incremental Conductance method was developed to overcome drawbacks associated with the P&O MPPT technique.

The typical flowchart for the Incremental Conductance method is shown in Figure 11 below.

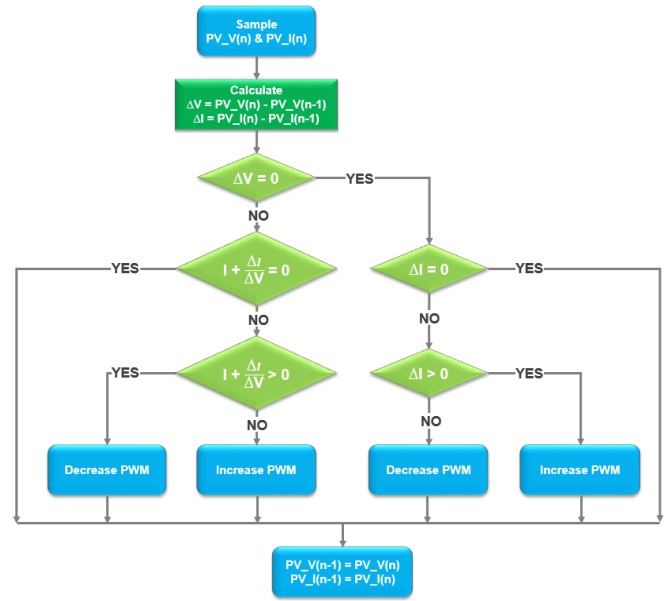

*Figure 11 - IC Functional Diagram*

The method computes the maximum power point by comparing the incremental conductance to the instantaneous conductance of the PV array, and when these are the same the MPP has been reached.

The increased data processing required for the Incremental Conductance method results in higher processor capacity usage. This means computational delays can be introduced for the complex function.

As mentioned, a limitation with the P&O algorithm is due to the step increase introducing oscillations when operating at the MPP. The advantage of using the incremental conductance method is there are fewer oscillations when the MPP is reached, however, due to a variety of factors it is difficult to maintain the exact MPP and therefore oscillations are still generated [32]. The Incremental Conductance method is more stable than the P&O method, at the expense of a delay in tracking capability and increased complexity [32].

## VI. MPPT Simulation

To examine the performance of each MPPT method simulation was conducted using Matlab Simulink. An irradiance profile was constructed to mirror a typical daily pattern.

The MPPT algorithms were developed in Matlab in accordance with Figure 10 and Figure 11. The results from the MPPT simulations are captured in Figure 12 and Figure 13 as well as Table 1.

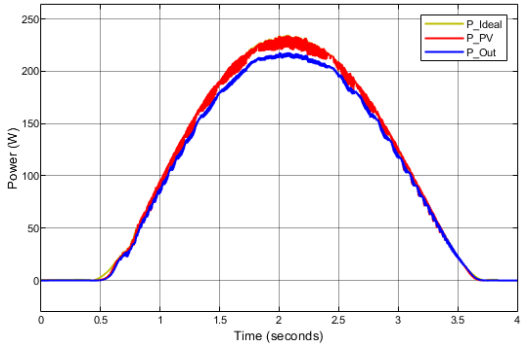

*Figure 12 - P&O Simulation Power Output Profile*

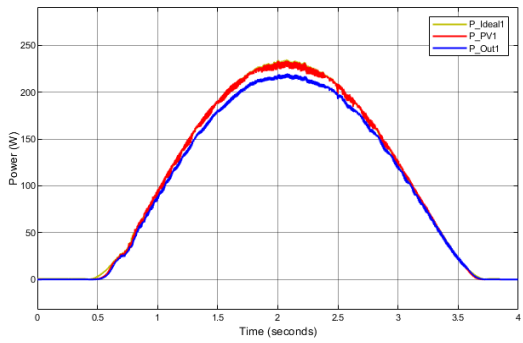

*Figure 13 - IC Simulation Output Power Profile*

| Parameter | Value |
| --- | --- |
| P&O RMS Power | 145.7 Prms |
| IC RMS Power | 146.7 Prms |
| P&O Efficiency | 93.0 % |
| IC Mean Efficiency | 93.6 % |
| P&O Converter Efficiency | 89.6 % |
| IC Converter Efficiency | 90.2 % |

*Table 1 – Simulation Output*

## VII. Hardware Implementation

The PCB was designed using EasyEDA which is online PCB design platform. To streamline the design and testing process, the micro-inverter was constructed using three PCB modules. There was a module designed for each of the following:

- DC-DC Boost Converter
- DC-AC Inverter
- LCL – Filter

### A. Gate Driver Design

The TLP351 photocoupler operates using a 10-30V supply voltage at Vcc (PIN 8). For this research, 12V was selected as an appropriate supply voltage. The 12V supply was connected to the TLP351 using an isolated DC-DC power supply. The chosen isolator was the TMH1212S which is a miniature device used to regulate the output to 12V. This ensured the consistent supply of power to each TLP351 photocoupler device. the gate driver for the MOSFET switches is shown in Figure 14.

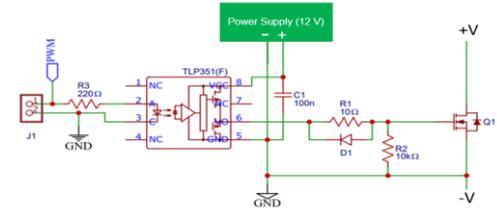

*Figure 14 - MOSFET Gate Driver Design*

### B. DC-DC Boost Converter

The circuit consists of the gate driver described above as well as the isolated power supply. The components chosen were rated according to the value, current and voltage related to each component from simulation completed using Matlab Simulink.

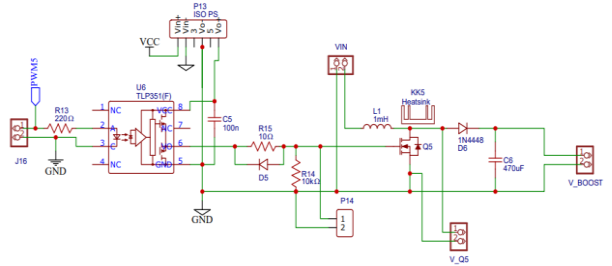

*Figure 15 - DC-DC Boost Converter Circuit Design*

### C. DC-AC Inverter

The inverter was designed using the boost converter PCB as a foundation. The gate driver configuration from the boost converter was replicated 4 times for the 4 MOSFETs required for the full-bridge inverter. The final circuit is shown in Figure 16.

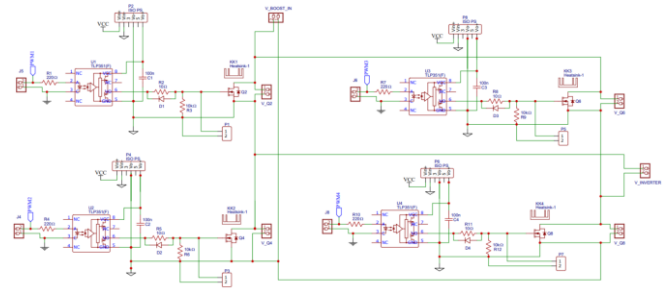

*Figure 16 - DC-AC Inverter Circuit Design*

### D. Control

The PSoC CY8CKIT 5LP Development Kit provides a control interface to the micro-inverter. The PSoC control interface was implemented using PSoC Creator which is an Integrated Design Environment (IDE) enabling firmware design. For this project the control mechanism uses PWM signals to control the boost converter and DC-AC inverter.

### E. Final Assembly

The final test hardware arrangement along with the LCL Filter module is shown in Figure 17 below. The final Bill of Materials (BoM) was £223, which included some spare components for replacements of faulty or damaged items.

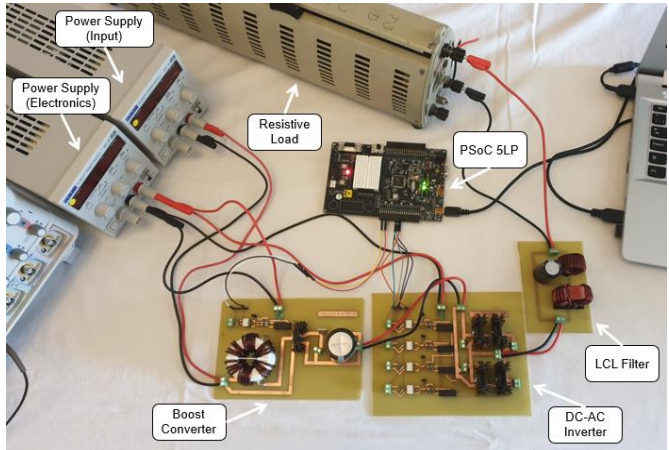

*Figure 17 - Hardware Assembly Test Arrangement*

### F. Hardware Output

The waveform in Figure 18 displays the output for the 10V inverter input voltage and 150Ω resistive load confirms the operation of the LCL Filter. The peak voltage $V_{pk}$ is 9.52V, therefore the voltage drop across the converter is 0.48V. The voltage drop could be reduced through choosing higher performance components for the MOSFETs, inductors and capacitor.

Additionally, the waveform does show some notches or inconsistencies in the sinusoidal output. These are expected to be a result of the non-ideal MOSFET switching which probed further analysis.

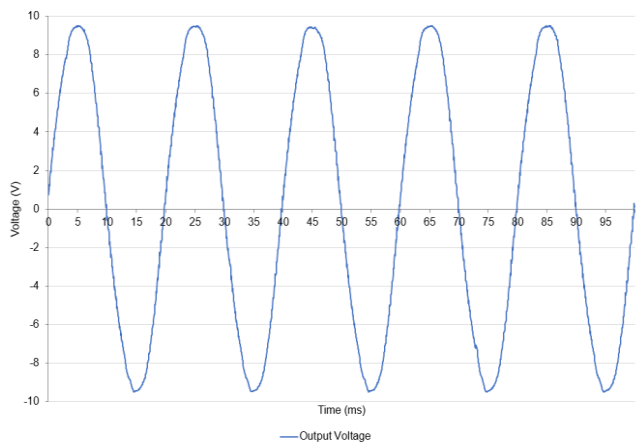

*Figure 18 - Output Voltage with 150Ω Load*

The waveform properties as measured by the oscilloscope are given Table 2.

| Parameter | Value |
|---|---|
| Output Frequency, $f_{out}$ | 49.94Hz |
| RMS Voltage, $V_{RMS}$ | 6.7335V |
| Maximum Voltage, $V_{pk}$ | 9.5226V |
| Peak-to-Peak Voltage, $V_{pk-pk}$ | 19.026V |

*Table 2 - Hardware Testing Output*

Further investigation into voltage drops realised a non-ideal switching pattern from the MOSFET switching devices. Figure 19 below shows significant overshoot upon the switch transitions for two series MOSFETs of the inverter module.

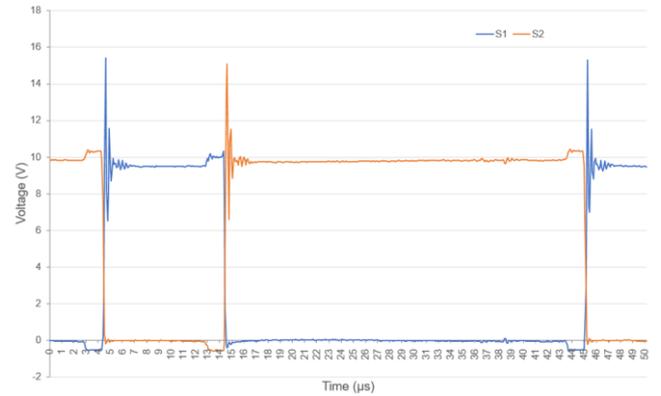

*Figure 19 - MOSFET S1 and S2 Switching Period*

The switching transient is due to the parasitic inductance and capacitance in the PCB routing, components, and wiring. The non-ideal switching of the MOSFETs will increase the switching losses as well as introducing harmonic content into the output waveform, reducing the inverter efficiency. To suppress the switching transients, snubber circuits are commonly used across the MOSFET to suppress the voltage transients and dampen the effects of ringing. Additionally, the amplitude of the surge voltage can be reduced by increasing the gate resistance.

### VIII. Conclusion

This paper has presented an introduction to the connection interfaces and characteristics required for grid connected inverter systems. An overview has been provided detailing popular topologies presented with key design features and limitations. Simulation and analysis of two popular MPPT techniques concludes the Incremental Conductance method has superior efficiency results due to reduced oscillation around the maximum power point. The CY8CKIT-050 PSoC 5LP Development Kit was used to provide the PWM signals to the boost converter and DC-AC inverter PCB modules. The testing was done using a 10V input supply which yielded a 6.74Vrms waveform with a peak of 9.56V. The voltage drop is due to the components being non-ideal as well as losses during the switching of the MOSFETs.